\documentclass[twocolumn]{autart}    

\usepackage{graphicx}          
\usepackage{natbib}          
\usepackage{amsmath}     
\usepackage{tikz}               
\usepackage{amssymb}
\usepackage{booktabs}

\newtheorem{definition}{Definition}{\bf}{}
\newtheorem{remark}{Remark}{\bf}{}
{\bf}{}
\newtheorem{theorem}{Theorem}{\bf}{}
\newtheorem{assumption}{Assumption}{\bf}{}
\newtheorem{corollary}{Corollary}{\bf}{}
{\bf}{}

\begin{document}

\begin{frontmatter}

\title{Bio-inspired Evolutionary Game Dynamics on Complex Networks under Uncertain Cross-inhibitory Signals \thanksref{footnoteinfo}} 

\thanks[footnoteinfo]{A short version of this work has appeared as invited paper in \cite{SB17}.}

\author[First]{Leonardo Stella}\ead{lstella1@sheffield.ac.uk} $\quad$
\author[Second]{Dario Bauso} \ead{d.bauso@sheffield.ac.uk}

\address[First]{Department of Automatic Control and Systems Engineering,  University of Sheffield, Mappin St. Sheffield, S1 3JD, UK.}
\address[Second]{Department of Automatic Control and Systems Engineering,  University of Sheffield, Mappin St. Sheffield, S1 3JD, United Kingdom, and Dip. di Ing. Chimica, Gestionale, Informatica, Meccanica,  Universit\`a di Palermo, 90128 Palermo, IT.}


\begin{abstract}                          
Given a large population of players, each player has three possible choices between option \emph{1} or \emph{2} or no option. The two options are equally favorable and the population has to reach consensus on one of the two options quickly and in a distributed way. The more popular an option is, the more likely it is to be chosen by uncommitted players. Uncommitted players can be attracted by those committed to any of the other two options through a cross-inhibitory signal. This model originates in the context of honeybees swarms, and we generalize it to  duopolistic competition and opinion dynamics. The contributions of this work include (1) the formulation of an evolutionary game model to explain the behavioral traits of the honeybees, (2) the study of the individuals and collective behavior including equilibrium points and stability, (3) the extension of the results to the case of structured environment via  complex network theory, (4) the analysis of the impact of the connectivity on \emph{consensus}, and (5) the study of absolute stability for the collective system under time-varying and uncertain cross-inhibitory parameter.
\end{abstract}

\end{frontmatter}


\section{Introduction}
We consider a large population of players who can choose option \emph{1}, option \emph{2} or no option (uncommitted state).  The two options are equally favorable and the population has to reach consensus on one of the two options quickly and in a distributed way. 
 Players i) benefit from choosing the more popular option, ii)  they can recruit uncommitted players, and iii) they can send cross-inhibitory signals to players committed to a different option.
%

\noindent \textbf{Highlights of contributions.} First, we provide an interpretation as game dynamics by modelling the evolution of the frequency of each strategy. We also introduce a new notion of game dynamics, which we call \emph{expected gain pairwise comparison}, according to  which the players change strategy with a probability that is proportional to the expected gain. We also extend the model to duopolistic competition and opinion dynamics. A second contribution is the analysis of stability of the individuals behaviors. Our analysis  shows that, if the cross-inhibitory parameter exceeds a threshold, which we calculate explicitly, players reach consensus on one of the two options. Otherwise they distribute uniformly across the two options at the equilibrium. 
As third contribution, we reframe the above results in the case of structured environment. The structure of the environment is captured by a complex network, with given degree distribution. The nodes  are the players and the degree of a node represents its connectivity. This allows us to study the role of heterogeneity. The following is a list of additional results with respect to the conference paper, see \cite{SB17}. First, we provide a convergence analysis as a function of the connectivity. Then, we prove that higher connectivity increases the number of uncommitted players. Last, we prove absolute stability under time-varying uncertain cross-inhibitory parameter.

\noindent \textbf{Related literature}. The proposed model originates in the context of  a swarm of honeybees, see \cite{Britton02}, and \cite{Marshall}. The analogy with duopolistic competition in marketing has been inspired by \cite{Bressan10}, and the link to opinion dynamics in social networks by  \cite{Hegselmann02}. Although the role of cross-inhibitory parameter was studied in \cite{Marshall}, here we stress a different perspective based on the Lyapunov's direct method for stability analysis and control design. Evolutionary dynamics in structured environment is discussed in \cite{Tan14}, \cite{PC08}, \cite{Ranjbar14}. A game perspective in collective decision making is provided in \cite{Sal15}. Consensus and games are studied in \cite{Yin12}.

This paper is organized as follows. In Section~\ref{sec:gm}, we describe the game.  In Section~\ref{sec:ex}, we discuss applications. In Sections~\ref{sec:ue} and~\ref{sec:se}, we consider unstructured and structured environments, respectively.  In Section~\ref{sec:asym}, we  study  the asymmetric case. In Section~\ref{sec:circle}, we study absolute stability under uncertain and time-varying cross-inhibitory signal. In Section~\ref{sec:ns}, we provide numerical analysis. In Section~\ref{sec:conc}, we provide conclusions and  future works. 

\section{Game Dynamics}\label{sec:gm}
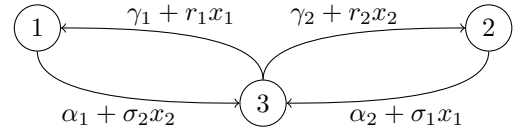
\begin{figure}[h]
\begin{centering}
      \begin{tikzpicture}
      \path 	(0,0) node[circle,draw](x) {$3$}
      		(-3,1) node[circle,draw] (z) {$1$}
		(3,1) node[circle,draw](y) {$2$};
	\draw[->,black] (x) .. controls +(up:1cm) and +(right:1cm) .. node[above] {$\gamma_1+r_1x_1$} (z);
	\draw[->,black] (z) .. controls +(down:1cm) and +(left:1cm) .. node[below] {$\alpha_1+\sigma_2x_2$} (x);
	\draw[->,black] (x) .. controls +(up:1cm) and +(left:1cm) .. node[above] {$\gamma_2+r_2x_2$} (y);
	\draw[->,black] (y) .. controls +(down:1cm) and +(right:1cm) .. node[below] {$\alpha_2+\sigma_1x_1$} (x);
      \end{tikzpicture}
\caption{Markov chain relating to the game dynamics.}
\label{markov}
\end{centering}
\end{figure}

Given a large population of players, each player chooses within a set of three pure strategies. Let us denote the frequency of strategy $i$, namely the portion of the population who has selected that strategy, by $x_i \in \mathbb{R}_0^+$, $\sum_{i=1}^3x_i=1$, for $i=1,2,3$. Let $A=(a_{ij})$ be the payoff matrix defined as follows:
\begin{equation} \label{eq:pm}
A = \left( \begin{array}{ccc}
r_1 & -\sigma_2 & 0 \\
-\sigma_1 & r_2 & 0 \\
0 & 0 & 0 \end{array} \right).
\end{equation}
The non-zero entries of matrix $A$ simulate a \emph{coordination game}, whereby the row player benefits from matching the column player's strategy. The row player earns $r_1$ and $r_2$ dollars for matching strategy $1$ or $2$, otherwise he loses $\sigma_1$ or $\sigma_2$ if playing strategy $2$ or $1$ while the column player plays the other strategy. Uncommitted players do not gain nor loose anything in random-matching with opponents. The above matrix models a \emph{crowd-seeking} scenario where the benefit of choosing a strategy between $1$ and $2$ depends on the frequency of that strategy. In addition, before choosing strategy $1$ or $2$, players must be in an uncommitted state, namely in strategy $3$.  
The evolution of the frequencies of each strategy  is in accordance with the following game dynamics  which links to the notion of  innovative dynamics as in \cite{Hofbauer11}. Let $\rho_{ij}$ be the transition rate from $i$ to~$j$:
\begin{equation} \label{eq:pw}
\dot{x}_i = \sum_j x_j\rho_{ji} - x_i\sum_j\rho_{ij}.
\end{equation}
The following is the definition of \textit{expected gain comparison} given $x$ for our game dynamics, which constitutes the first contribution of this paper.
\begin{definition} \textbf{(Expected gain comparison)} 
Given a payoff matrix $A=(a_{ij}) \in \mathbb R^{n \times n}$, by changing from strategy $j$ to  $i$ the  expected gain pairwise payoff comparison is defined as
\begin{equation} \label{eq:eg}
E_{ji}(x) = \sum_{k=1}^n (a_{ik} - a_{jk})_+ x_k + b_{ji},
\end{equation}
where $(a_{ik} - a_{jk})_+$ denotes the positive part of $a_{ik} - a_{jk}$, and $b_{ji}$ is an offset.
\end{definition}
The above definition models the expected revenue obtained by considering the probability only of a payoff increase and ignoring payoff decreases in correspondence to a unilateral change of strategy.\\
For the payoff matrix in (\ref{eq:pm}), we then have $\rho_{31} = r_1 x_1 +\gamma_1$, $\rho_{13} = \sigma_2 x_2 +\alpha_1$, $\rho_{32} = r_2 x_2 +\gamma_2$ and $\rho_{23} = \sigma_1 x_1 +\alpha_2$, where the offset $b_{ij}$ is $\gamma_1$, $\gamma_2$, $\alpha_1$ or $\alpha_2$ in each specific case.
By substituting the previous equations in dynamics (\ref{eq:pw}) and using the conservation of mass law, for which it holds $\dot x_3=-\dot x_1-\dot x_2$, the  formulation of the system can be reduced to a two-dimensional system as follows:
\begin{equation} \label{eq:gfs}
\mbox{Unstructured} \, \left\{\begin{array}{lll}
\dot{x}_1 = x_3(rx_1+\gamma) - x_1(\alpha+\sigma x_2), \\
\dot{x}_2 = x_3(rx_2+\gamma) - x_2(\alpha+\sigma x_1), \\
\end{array}\right.
\end{equation}
where we take $\alpha_1=\alpha_2=:\alpha$, $\gamma_1=\gamma_2=:\gamma$ and $\sigma_1=\sigma_2=:\sigma$ (symmetric case). The above system is obtained in the case of \emph{unstructured} environment, i.e., it does not consider any interaction topology. Such a system, in the asymmetric case, where the parameters are different for the two options, admits the Markov chain representation displayed in Fig. \ref{markov}.

In the case of structured environment, let a complex network be given where $P(k)$ is the probability distribution of the node degrees. Also let $x_i^k$ be the portion of the population with $k$ connections (class $k$ in short) using strategy $i$, and let $\psi_k = \frac{k}{k_{max}}$ be the parameter capturing the connections of the players in the network. Furthermore, let  $\langle k \rangle$ be the mean value of $k$, and let $\theta_i:=\frac{1}{\langle k \rangle}\sum_k k P(k) x_i^k $ be the probability that a link randomly chosen will point to a player using strategy $i$. The counterpart of system (\ref{eq:gfs}) for every class $k \in \mathbb{Z}^+$ is: 
\begin{equation}\label{eq:psist}
\mbox{Structured}  \left\{\begin{array}{ll}
\dot{x}_1^k =  x_3^k(\psi_k r \theta_1+\gamma) - x_1^k(\alpha+ \psi_k\sigma \theta_2^k), \\
\dot{x}_2^k = x_3^k(\psi_kr\theta_2+\gamma) - x_2^k(\alpha+ \psi_k\sigma \theta_1). 
\end{array}\right.
\end{equation}
We can view the  above system as a microscopic model of the players in class $k$ parametrized by the macroscopic parameters $\theta_1$ and $\theta_2$. 

\section{Examples}\label{sec:ex}
This section discusses three examples of applications of the game model in (\ref{eq:gfs}), namely honeybees swarm, duopolistic competition and opinion dynamics.

\noindent
\textbf{Swarm of honeybees}. System (\ref{eq:gfs}) was first developed in the context of honeybees swarms, see \cite{Marshall}. The swarm has to choose between two nest-boxes. The two options  have same value $r_1=r_2=:r$. Scout bees recruit uncommitted bees via a ``waggle dance''. The parameters $\sigma_1=\sigma_2=:\sigma$ weight the strength of the cross-inhibitory signals. We can interpret $x_1$ as the portion of swarm selecting option $1$, $x_2$ the portion of swarm selecting $2$ and $x_3$ the portion of swarm in the uncommitted state $3$. Transitions from option $3$ to  $1$ involve a $\gamma_1$ amount of independent bees that choose $1$ spontaneously and  a quantity $\rho_1x_1$  of bees attracted by those who are already in $1$.  On the other hand, consider the case where bees move from strategy $1$ to $3$: $\alpha_1$ are those that spontaneously abandon their commitment to strategy $1$ and $\sigma_2 x_2$ takes into account the cross-inhibitory signal sent from bees using option~$2$. 
\noindent
\textbf{Duopolistic competition in marketing}. System (\ref{eq:gfs}) provides an alternative model of duopolistic competition in marketing, see e.g. Example 9, p. 27 in \cite{Bressan10}. The classical scenario captured by the well-known  \emph{Lanchester}  model is as follows. Two manufacturers produce the same product in the same market. The variables $x_i$ represent the market share of the manufacturer $i$ at time $t$. The cross-inhibitory signal and the ``waggle dance'' term describe different advertising efforts, which may enter the problem as parameters or controlled inputs in the analysis or design of the advertising campaign.  Thus system (\ref{eq:gfs}), likewise the \emph{Lanchester}  model, describes the evolution of the market share.  
In the case of structured environment, system (\ref{eq:psist}) captures the \emph{social influence} of the advertisement campaigns of both manufacturers. A stronger cross-inhibitory signal can be used to model the capability of reaching out to a larger number of potential clients. 

\noindent
\textbf{Opinion dynamics}. Consider a  population of individuals, each of which can prefer to  vote \emph{left} or \emph{right}, see \cite{Hegselmann02}.  This is represented by the Markov chain depicted in Fig.~\ref{markov} where nodes $1$ and $2$ represent the \emph{left} and \emph{right}. The distribution of individuals in each state is subject to transitions from one state to the other.  
Persuaders who campaign for the \emph{left} can influence the transitions from the \emph{right} to the \emph{uncommitted state} in a similar way honeybees use cross-inhibitory signals.  At the same time \emph{uncommitted} individuals select \emph{left} or \emph{right} proportionally to the level of popularity of the two options.  
In the case of structured environment, system (\ref{eq:psist}) captures the \emph{social influence} of each individual. In other words, the cross-inhibitory signal is stronger for those individuals who have more connections. 

%
%
%
%
%
%
%
%

\section{Unstructured environment}\label{sec:ue}
In this section, we study  stability under unstructured environment and symmetric cross-inhibitory parameters.

\begin{theorem}\label{th1}
Given $T>0$ and an initial state $\mathbf{x}_0$, the equilibrium points of game dynamics (\ref{eq:gfs}) are:
\begin{itemize}
\item \textbf{Case 1}. When $x_1 = x_2$,
\begin{equation} \label{eq:eqcase1}\nonumber
\begin{array}{lll}
x^*= (x_1^*,x_2^*,x_3^*) = \Big(\frac{(r-2\gamma-\alpha) + \sqrt{(r-2\gamma-\alpha)^2 + 4\gamma(2r +\sigma)}}{2(2r +\sigma)}, \\
\qquad \qquad \frac{(r-2\gamma-\alpha) + \sqrt{(r-2\gamma-\alpha)^2 + 4\gamma(2r +\sigma)}}{2(2r +\sigma)}, \\
\qquad \qquad 1-\frac{(r-2\gamma-\alpha) + \sqrt{(r-2\gamma-\alpha)^2 + 4\gamma(2r +\sigma)}}{2r +\sigma}\Big).
\end{array}
\end{equation}
\item \textbf{Case 2}. When $x_3 = \alpha/r$,
\begin{equation} \label{eq:eqcase2}\nonumber
\begin{array}{lll}x^*= (x_1^*,x_2^*,x_3^*) = \Big(
\frac{1-\frac{\alpha}{r} \pm \sqrt{(1-\frac{\alpha}{r})^2 + \frac{4 \alpha \gamma}{\sigma r}}}{2}, \\
\qquad \qquad 1- \frac{1-\frac{\alpha}{r} \pm \sqrt{(1-\frac{\alpha}{r})^2 + \frac{4 \alpha \gamma}{\sigma r}}}{2}- \frac{\alpha}{r} , \frac{\alpha}{r}\Big).
\end{array}
\end{equation}
\item \textbf{Case 3}. When $x_1 = x_2$ and $x_3 = \alpha/r$,
\begin{equation} \label{eq:eqcase3}\nonumber
\begin{array}{lll}
x^*= (x_1^*,x_2^*,x_3^*) = \Big(
\sqrt{\frac{\alpha \gamma}{r \sigma}}, 
\sqrt{\frac{\alpha \gamma}{r \sigma}}, \frac{\alpha}{r}\Big)
=  \Big(
\frac{r-\alpha}{2r}, 
\frac{r-\alpha}{2r}, \frac{\alpha}{r}\Big).
\end{array}
\end{equation}
\end{itemize}
\end{theorem}
Cases $1$ and $3$ refer to equilibrium points that are symmetric, i.e. we have the same number of individuals committed to option $1$ and $2$. 

\begin{corollary}
Let $\alpha \rightarrow 0$, the equilibria converge to $(1,0,0)$ and $(0,1,0)$ in \textbf{Case 2} and to $(\frac{1}{2},\frac{1}{2},0)$ in \textbf{Case~3}.
\end{corollary}
Note that the equilibrium points $(1,0,0)$ and $(0,1,0)$ correspond to consensus to option 1 and 2 respectively, while $(\frac{1}{2},\frac{1}{2},0)$ means that players are uniformly distributed between the two options. 
These results will be used in the following sections, when we will consider a time-varying cross-inhibitory signal $\sigma(t)$, which is one of the novelties of this work.
The next result establishes local asymptotic stability of the symmetric equilibrium described in Case 1.
\begin{theorem}\label{th2}
Given $T>0$ and an initial state $\mathbf{x}_0$, the symmetric equilibrium point in Case 1 is locally asymptotically stable if and only if
\begin{equation} \label{eq:sigmath}
\sigma < \frac{4r\alpha \gamma}{(r-\alpha)^2}.
\end{equation}
\end{theorem}

\begin{remark}
In the special case where $\alpha=\frac{1}{r}$ and $\gamma=r$ our results are in accordance with the threshold value reported in equation (4) in \cite{Marshall}. 
\end{remark}

\section{Structured environment}\label{sec:se}
In this section we extend to the case of structured environment the results on equilibrium points and stability provided in the previous section. 
Let us consider game dynamics (\ref{eq:psist}) and  analyze the mean-field response obtained for a given class of players assuming that the distribution of the rest of the population is fixed. 
 From  $x_1^k + x_2^k + x_3^k=1$,  game dynamics (\ref{eq:psist}) becomes
\begin{equation}\label{eq:psist11111}
\begin{array}{ll}
\dot{x}_1^k = (1 - x_1^k - x_2^k)(\psi_kr \theta_1+\gamma) - x_1^k(\alpha+ \psi_k\sigma \theta_2), \\
\dot{x}_2^k = (1 - x_1^k - x_2^k)(\psi_kr \theta_2+\gamma) - x_2^k(\alpha+ \psi_k\sigma \theta_2).
\end{array}
\end{equation}

We can rewrite the above system in matrix form and,
under the assumption that $\theta_1=\theta_2=:\theta$, we have
\begin{equation} \label{eq:mf2}
\begin{array}{ll}
\left[ \begin{array}{c}
\dot x_1^k \\
\dot x_2^k \end{array} \right] = 
\overbrace{\left[ \begin{array}{cc}
- (r + \sigma)  \psi_k  \theta  - \alpha - \gamma  & 
- \psi_k r \theta - \gamma \\
- \psi_k r \theta - \gamma  & - (r + \sigma)  \psi_k  \theta  - \alpha - \gamma  
\end{array} \right]}^{A_k(\theta)} \\
\qquad \qquad \cdot 
\left[ \begin{array}{c}
x_1^k \\
x_2^k \end{array} \right] +
\underbrace{\left[ \begin{array}{c}
\psi_k r \theta + \gamma  \\
\psi_k r \theta + \gamma \end{array} \right]}_{c_k(\theta)}.
\end{array}
\end{equation}


\begin{theorem}\label{th3}
Given $T>0$ and an initial state $\mathbf{x}^k_0$, for all classes $k$, system (\ref{eq:mf2}) is locally asymptotically stable and convergence is faster with increasing connectivity $\psi_k$. Furthermore, in the cases of no connectivity $\psi_k = 0$ and full connectivity $\psi_k = 1$, system (\ref{eq:mf2}) has eigenvalues 
\begin{equation}\label{eq:labdas}\nonumber
\lambda_{1,2} = \left\{ \begin{array}{ll}
 (-\alpha -2\gamma, -\alpha), \, for \, \psi_k = 0, \\
 (-(2r + \sigma)\theta -\alpha -2\gamma, -\sigma\theta -\alpha), \, for \, \psi_k = 1.
\end{array} \right.
\end{equation}
\end{theorem}

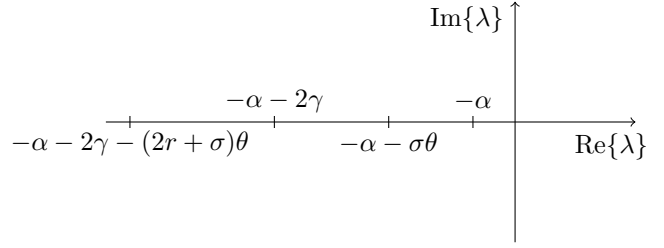
\begin{figure} [h]
\centering
\def\svgwidth{.8\columnwidth}
\begin{tikzpicture}
  \begin{scope}[shift={(0,0)},scale=0.8]
  \draw [->,] (-4.4,0) -- (4.4,0);
  \draw [->,] (2.4,-2) -- (2.4,2);
  \node at (4,-0.4) {Re$\{\lambda\}$};
  \node at (1.6,1.7) {Im$\{\lambda\}$};
  \draw (-1.6,-0.1) -- (-1.6,0.1);
  \node at (-1.6,0.35) {$-\alpha -2\gamma$};
  \draw (1.7,-0.1) -- (1.7,0.1);
  \node at (1.7,0.35) {$-\alpha$};
  \draw (-4,-0.1) -- (-4,0.1);
  \node at (-4,-0.35) {$-\alpha-2\gamma-(2r + \sigma)\theta$};
  \draw (0.3,-0.1) -- (0.3,0.1);
  \node at (0.3,-0.35) {$-\alpha - \sigma\theta$};
  \end{scope}
      \end{tikzpicture}
      \caption{Change of the eigenvalues for system (\ref{eq:mf2}).}
      \label{fig:eig}
\end{figure}

From Fig. \ref{fig:eig}, we can see that the connectivity shifts the eigenvalue further away from the origin (the ones for the first case are labelled above the $x$-axis, while the ones for the second case are below). Thus, higher connectivity speeds up convergence. 

\begin{theorem}\label{th4}
Let $T>0$ and an initial state $\mathbf{x}^k_0$, for class~$k$, the equilibrium points are 
\begin{equation}\label{eq:compform}
\begin{array}{lll}
\left[ \begin{array}{c}
 \hat x_1^k \\
\hat x_2^k \end{array} \right] = - A^{-1}_k(\theta)  c_k(\theta).
\end{array}
\end{equation}
Furthermore, at the equilibrium, the distribution $x_3$ of uncommitted players increases with connectivity $\psi_k$.
\end{theorem}

\begin{remark}
The physical interpretation of the above result is that by increasing the connectivity of the network we bring more uncertainty into the collective decision making process. This reflects in an increase of the percentage of uncommitted players at steady-state. 
\end{remark}
Let us now develop a model combining a macroscopic and microscopic dynamics. By averaging on both sides of (\ref{eq:mf2}) using $\frac{1}{\langle k \rangle}\sum_k kP(k)$ we have the following macroscopic model:
\begin{equation}\label{eq:psi2}
\left\{\begin{array}{ll}
\dot{\theta}_1 = \frac{r\theta_1}{k_{max}}\Big(\frac{V(k)}{\langle k \rangle} - \Psi_1 - \Psi_2\Big) - \frac{\sigma \theta_2}{k_{max}}\Psi_1 \\
\qquad \qquad - \theta_1\alpha + \gamma -\theta_1\gamma - \theta_2\gamma, \\
\dot{\theta}_2 = \frac{r\theta_2}{k_{max}}(V(k)/\langle k \rangle - \Psi_1 - \Psi_2) - \frac{\sigma \theta_1}{k_{max}}\Psi_2 \\
\qquad \qquad - \theta_2\alpha + \gamma -\theta_1\gamma - \theta_2\gamma,
\end{array}\right.
\end{equation}
where $V(k) = \sum_k k^2P(k)x^k$ and $\Psi = \frac{1}{\langle k \rangle}  \sum_k k^2P(k)x^k$.
\begin{theorem}\label{thsig}
Given $T>0$ and an initial state $\mathbf{x}_0$, the symmetric equilibrium point in the case of structured environment is locally asymptotically stable if and only if
\begin{equation} \label{eq:psi5}
\sigma < 2r - \frac{rV(k)}{\langle k \rangle \Psi} + \frac{\alpha k_{max}}{\Psi}.
\end{equation}
\end{theorem}
The above threshold for the cross-inhibitory signal generalizes (\ref{eq:sigmath}) in the case of structured environment. When $k = k_{max}$, i.e. in the case of fully connected network, the threshold in (\ref{eq:psi5}) coincides with (\ref{eq:sigmath}).

\section{The asymmetric case}\label{sec:asym}
\begin{figure}[h]
\begin{centering}
      \begin{tikzpicture}
      \path 	(0,0) node[circle,draw](x) {$3$}
      		(-3,1) node[circle,draw] (z) {$1$}
		(3,1) node[circle,draw](y) {$2$};
	\draw[->,black] (x) .. controls +(up:1cm) and +(right:1cm) .. node[above] {$\gamma_1$} (z);
	\draw[->,black] (x) .. controls +(up:1cm) and +(left:1cm) .. node[above] {$\gamma_2$} (y);
	\draw[->,black] (y) .. controls +(down:1cm) and +(right:1cm) .. node[below] {$\sigma\Theta_1$} (x);
      \end{tikzpicture}
\caption{Markov chain for asymmetric case.}
\label{markov4}
\end{centering}
\end{figure}
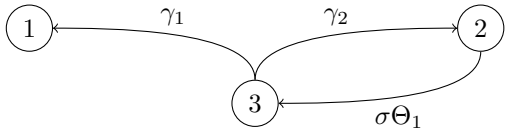

In the asymmetric case we consider only the cross-inhibitory signal sent from players in $1$ to players in $2$ and the  spontaneous migration from $3$ to $1$ and $2$ with rate $\gamma_1$ and $\gamma_2$ respectively. 
The resulting model is
\begin{equation} \label{eq:sircnn}
\begin{array}{ll}
\dot{x}_1 = \gamma_1 x_3,\\
\dot{x}_2 = -\sigma  x_2 x_1 + \gamma_2 x_3, \\
\dot{x}_3 = -\gamma_1 x_3 - \gamma_2 x_3 + \sigma x_2 x_1.\\
\end{array}
\end{equation}
The above dynamics share striking similarities with  the susceptible-infected-removed (SIR) model. Actually, $x_1$, $x_2$ and $x_3$ can be viewed as the percentage of susceptible, infected and recovered agents, respectively. Parameter $\gamma_1$ is the rate at which individuals decay into the recovered class and parameter $\sigma$ is the rate at which the infection is spread among the population.
The counterpart of (\ref{eq:sircnn}) in the case of heterogeneous connectivity is
\begin{equation} \label{eq:sircn}
\begin{array}{ll}
\dot{x}_1^k = \gamma_1 x_3,\\
\dot{x}_2^k = -\sigma \psi_k x_2^k \Theta_1 + \gamma_2 x_3, \\
\dot{x}_3^k = -\gamma_1 x_3 - \gamma_2 x_3 + \sigma \psi_k x_2^k \Theta_1, \\
\end{array}
\end{equation}
where the coefficients have the known meaning and $\Theta_1$ is a function of time that captures the probability that any given link points to a player in $1$. 
System (\ref{eq:sircn}) can be represented by the  Markov chain in Fig. \ref{markov4}.

Furthermore we define function $\Theta_1$ as: 
\begin{equation} \label{eq:theta}
\Theta_1 = \frac{\sum_k kP(k)x_1^k}{\sum_j jP(j)} = \frac{\sum_k kP(k)x_1^k}{\langle k \rangle},
\end{equation}
and its first derivative  $\Psi$ as:
\begin{equation} \label{eq:psi}
\dot \Theta_1 = \frac{\sum_k kP(k)\dot x_1^k}{\langle k \rangle} = \frac{\sum_k kP(k)\gamma_1 x_3}{\langle k \rangle} := \Psi.
\end{equation}
Consider the second derivative of $x_2^k$:
\begin{equation} \label{eq:s2}
\ddot x_2^k = -\sigma \dot x_2^k \Theta_1 - \sigma x_2^k \dot \Theta_1 +\gamma \dot x_3^k.
\end{equation}
The above second-order differential equation corresponds to the following bidimensional first-order system: 
\begin{equation} \label{eq:s2sys}
\left[ \begin{array}{c}
\dot x_2^k \\
\ddot x_2^k \end{array} \right] = 
\left[ \begin{array}{cc}
0 & 1\\
-\sigma \Psi \frac{k}{k_{max}}& -\sigma \Theta_1 \frac{k}{k_{max}} \end{array} \right]
\left[ \begin{array}{c}
x_2^k \\
\dot x_2^k \end{array} \right] +
\left[ \begin{array}{c}
\gamma x_3^k \\
\gamma \dot x_3^k \end{array} \right].
\end{equation}

The above  system shares similarities with a mass-spring-damper, where $\Theta$ plays the role of viscous term, while the eigenvalues determine the amplitude of the oscillations. Implication of such a mechanical analogy will be highlighted and  discussed further in the section on  the numerical analysis.

\section{Uncertain cross-inhibitory coefficient}\label{sec:circle}
\begin{figure}[h]
\centering
\begin{tikzpicture}
\draw (0.2,-0.25) rectangle (2,0.75);\node at (1.1,0.25) {
 $G(s)$};
\draw (0.2,-2.0) rectangle (0.2+1.8,-2.0+1.0);\node at (.2+.9,-2+.5) {
$\psi(t)$};
\draw (0.2,-1.5) -- (-1.0,-1.5);
\node at (-1.0,-1.8) {$f(y)$};
\draw [->,  thick]  (-1.0,-1.5) -- (-1.0,-0.15);
\node at (-1.3,-0.4) {$-$};
 \path
(-1.0,0.2) node[circle,draw](x) {$+$};
\draw [->,  thick]  (-2.0,0.2) -- (-1.35,0.2);
\node at (-1.7,0.5) {$r(t)$};
\draw [->,  thick] (-0.65,0.2) -- (0.2,0.2);
\node at (-0.2,0.5) {$e(t)$};
\draw [->,  thick] (0.5,1.4) -- (0.5,0.75);
\node at (0.7,1.05) {$k$};
\draw (2.8,0.2) -- (2.8,-1.5); \draw [->, thick] (2.8,-1.5) -- (2.0,-1.5);
\draw [->,  thick] (2,0.2) -- (3.8,0.2);
\node at (3.4,0.5) {$y(t)$};
\end{tikzpicture}
\caption{Feedback scheme used to isolate the nonlinearity of the asymmetric  (\ref{eq:sircnn}) and symmetric systems (\ref{eq:gfs}).}
\label{fig:tfa}
\end{figure}
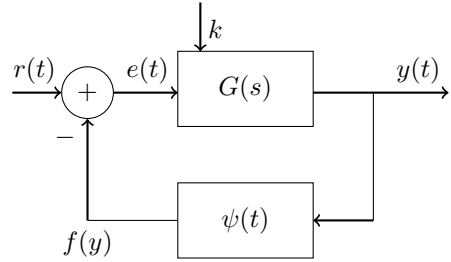

In this section, we show that stability properties are not compromised even if the cross-inhibitory coefficient $\sigma$ is uncertain and changes with time within a pre-specified interval. To do this, we first isolate the nonlinearity related to the cross-inhibitory signal in the feedback loop and prove absolute stability using the Kalman-Yakubovich-Popov lemma, see Chapter 10.1 in \cite{Khalil02}. The feedback scheme used in this section is depicted in Fig. \ref{fig:tfa}.

Now, the system described by the following set of equations is considered: 
\begin{equation} \label{eq:cccs}
\begin{array}{lll}
\dot{x}_1 = (1 - x_1 - x_2)(rx_1 + \gamma) - x_2 ( \sigma x_1 + \alpha), \\
\dot{x}_2 = (1 - x_1 - x_2)(rx_2 + \gamma) - x_1 ( \sigma x_2 + \alpha). \\
\end{array}
\end{equation}
In the following assumption we introduce the sector nonlinerarities.
\begin{assumption}\label{as1}
Let the cross-inhibitory coefficient $\sigma$ be in $[0, \tilde k]$. 
\end{assumption}
We assume, for simplicity, that $x_1 = x_2$. Thus, we can write
\begin{equation} \label{eq:cccs2}
\begin{array}{lll}
\dot{x} = (1 - 2x)(rx + \gamma) - x ( \sigma x + \alpha).
\end{array}
\end{equation}
The linearized version of (\ref{eq:cccs}) is
\begin{equation} \label{eq:cccsc}
\begin{scriptsize}
\left[\begin{array}{c}
\dot x_1 \\
\dot x_2 \end{array} \right]
=\underbrace{\left[ \begin{array}{cc}
r-3rx-\gamma-\alpha \quad & -rx-\gamma \\
-rx-\gamma \quad & r-3rx-\gamma-\alpha  \end{array} \right] }_{A}
\left[\begin{array}{c}
x_1 \\
x_2 \end{array} \right].
\end{scriptsize}
\end{equation} 
Building on the Kalman-Yakubovich-Popov lemma, absolute stability is linked to strictly positive realness of $Z(s) = \mathbb{I} + KG(s)$ where $K=\tilde k \mathbf 1 \mathbf 1^T \in \mathbb R^{2 \times 2}$ and $G(s)$ is the transfer function of system (\ref{eq:cccsc}), yet to be calculated. Before addressing absolute stability, we first investigate conditions under which matrix $A$ is Hurwitz. To be Hurwitz, the trace of matrix $A$ must be negative, i.e. $Tr(A) = 2(r-3rx-\gamma-\alpha)<0$, and the determinant must be positive, i.e. $\Delta(A) = (r-3rx-\gamma-\alpha)^2 -  (-rx-\gamma)^2 > 0$. For the first condition, we can neglect the multiplier and have
$$Tr(A) = r-3rx-\gamma-\alpha \le r(1 - 3/2) -\gamma -\alpha,$$
where the equality holds from the condition $x_1=x_2=:x$ which implies, in turn, that $x$ can be at most 0.5. In the case where $x$ is sufficiently small, $\gamma$ and $\alpha$ can be set sufficiently large to guarantee the condition $Tr(A) <0$. For the condition on the determinant, we have
$$\Delta(A) = 3rx+\gamma+\alpha-r-rx-\gamma  = 2rx + \alpha - r > 0,$$
which is satisfied, when $x=0.5$, and is still true by choosing a proper $\alpha$ in all the other cases.
Now, we isolate the nonlinearities in $\psi$, and we set $B = C = \mathbb{I}$, where $\mathbb{I}$ denotes the identity matrix. Let us now obtain the transfer function associated with system (\ref{eq:cccsc}):
\begin{equation} \label{eq:cccscG}
G(s) = c^T[s\mathbb{I} - A]^{-1}b = \frac{1}{a^2-b^2}
\begin{scriptsize}
\left[ \begin{array}{cc}
a & -b \\
-b & a  \end{array} \right], 
\end{scriptsize}
\end{equation}
where $a = s+3rx+\gamma+\alpha-r$ and $b = rx+\gamma$. Then, for $Z(s)$ we obtain
\begin{equation} \label{eq:cccscZ}
\begin{scriptsize}
\begin{array}{lll}
Z(s) = \mathbb{I} + KG(s) = 
\left[ \begin{array}{cc}
1 & 0 \\
0 & 1  \end{array} \right] +
\left[ \begin{array}{cc}
\frac{ak - bk}{a^2-b^2} & \frac{-bk + ak}{a^2-b^2} \\
\frac{-bk + ak}{a^2-b^2} & \frac{ak - bk}{a^2-b^2}  \end{array} \right] 
\\ \\ = \left[ \begin{array}{cc}
1 + \frac{k}{a+b} & \frac{k}{a+b} \\
\frac{k}{a+b} & 1 + \frac{k}{a+b}  \end{array} \right] = 
 \frac{1}{s+\zeta}
\left[ \begin{array}{cc}
s+\zeta + k & k \\
k & s+\zeta + k  \end{array} \right],
\end{array}
\end{scriptsize}
\end{equation}
where $\zeta = 4rx+2\gamma+\alpha - r$. We are ready to establish the following result.

\begin{theorem}\label{th5}
Let system (\ref{eq:cccsc}) be given and assume that $A$ is Hurwitz. Furthermore, let us consider the sector nonlinearity as in Assumption \ref{as1}. Then, $Z(s)$ is strictly positive real and the system (\ref{eq:cccsc}) is absolutely stable. 
\end{theorem}

We can extend our robustness analysis to the asymmetric system described by the following set of equations: 
\begin{equation} \label{eq:ccas}
\begin{array}{lll}
\dot{x}_1 = \gamma_1(1 - x_1 - x_2), \\
\dot{x}_2 = - \sigma_1 x_1 x_2 +  \gamma_2(1 - x_1 - x_2). \\
\end{array}
\end{equation}
System (\ref{eq:ccas}) is the asymmetric version of (\ref{eq:gfs}), when $r$ and $\alpha$ are negligible. This system admits two equilibrium points, i.e. $x^* = (1,0),(0,1)$. As in the previous sections, by applying the Lyapunov linearisation method we study the stability of these equilibrium points. 
For the equilibrium point  $x^* = (1,0)$ and $x^* = (0,1)$, the Jacobian matrices are given by
\begin{equation} \label{eq:ccjac1} \nonumber
\begin{scriptsize}
J_{(1,0)}=\left[ \begin{array}{cc}
-\gamma_1 \quad & -\gamma_1 \\
-\gamma_2 \quad & - \gamma_2  -\sigma_1 \end{array} \right], \quad
J_{(0,1)}=\left[ \begin{array}{cc}
-\gamma_1 \quad & -\gamma_1 \\
-\gamma_2 - \sigma_1 \quad & - \gamma_2  \end{array} \right].
\end{scriptsize}
\end{equation}
For $J_{(1,0)}$ the trace is $T = -\gamma_1 - \gamma_2 -\sigma_1 < 0$ and the determinant is $\Delta = -\gamma_1 (- \gamma_2 - \sigma_1) - \gamma_1 \gamma_2 = \gamma_1 \sigma_1 > 0$, which means that  $x^* = (1,0)$ is stable. Analogously, for $J_{(0,1)}$,
the trace is $T = -\gamma_1 - \gamma_2 < 0$ and the determinant is $\Delta = \gamma_1 \gamma_2 - \gamma_1 (\gamma_2 + \sigma_1) = -\gamma_1 \sigma_1 < 0$, which means that  $x^* = (0,1)$ is a saddle. The corresponding bidimensional first-order system is
\begin{equation} \label{eq:ccas2}
\begin{array}{lll}
\left[ \begin{array}{c}
\dot x_1 \\
\dot x_2 \end{array} \right] = 
\left[ \begin{array}{cc}
-\gamma_1 & -\gamma_1 \\
-\gamma_2 & -\gamma_2 \end{array} \right]
\left[ \begin{array}{c}
x_1 \\
x_2 \end{array} \right] +
\left[ \begin{array}{c}
\gamma_1  \\
\gamma_2  \end{array} \right] -
\left[ \begin{array}{c}
0  \\
\sigma_1 \end{array} \right] x_1 x_2, \\ \\
y = \Big[0 \quad 1\Big] \left[ \begin{array}{c} x_1  \\ x_2 \end{array} \right].
\end{array}
\end{equation}
We denote the matrix of $\gamma_1,\gamma_2$ as matrix $A$, the constant vector $[\gamma_1 \quad \gamma_2]^T$ as $k$, the vector $[0 \quad \sigma_1]^T$ as $b$ and the vector $[0 \quad 1]$ as $c^T$. Here, we denote $A^T$ for the transpose of matrix $A$. For a first approximation, we will not consider vector $k$. The resulting calculations for the transfer function are:
\begin{equation} \label{eq:cctf}
\begin{array}{lll}
G(s) = c^T[s\mathbb{I} - A]^{-1}b = \frac{(s+\gamma_1)\sigma_1}{s(s+\gamma_1+\gamma_2)}.
\end{array}
\end{equation}
Now, we check whether the transfer function is positive real, to ensure stability of system (\ref{eq:ccas}). To be positive real, the following conditions must hold true: 

\noindent
\textbf{(1)} $G(s)$ is stable, i.e. no poles Re$\{s\}>0$. 

\noindent 
 \textbf{(2)} Re$\{G(j\omega)\}\geq 0$, i.e. $-\pi/2 \leq G(j\omega) \leq \pi/2$.

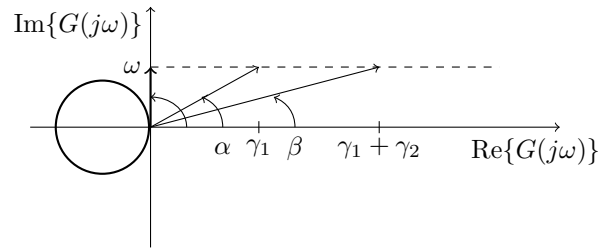
\begin{figure} [h]
\centering
\def\svgwidth{.8\columnwidth}
\begin{tikzpicture}
  \begin{scope}[shift={(0,0)},scale=0.8]
  \draw[thick] (-3.2,0) circle (0.775cm);
  \draw [->,] (-4.4,0) -- (4.4,0);
  \draw [->,] (-2.4,-2) -- (-2.4,2);
  \node at (4,-0.4) {Re$\{G(j\omega)\}$};
  \node at (-3.6,1.7) {Im$\{G(j\omega)\}$};
  \draw (-0.6,-0.1) -- (-0.6,0.1);
  \node at (-0.6,-0.35) {$\gamma_1$};
  \draw (1.4,-0.1) -- (1.4,0.1);
  \node at (1.4,-0.35) {$\gamma_1+\gamma_2$};
  \draw [->,] (-2.4,0) -- (-0.6,1);
  \draw [->,](-1.2,0) to[out=+90,in=-22.5] (-1.55,0.5);
  \node at (-1.2,-0.35) {$\alpha$};
  \draw [->,] (-2.4,0) -- (1.4,1);
  \draw [->,](0,0) to[out=+90,in=-22.5] (-0.35,0.5);
  \node at (0,-0.35) {$\beta$};
  \draw [->, thick] (-2.4,0) -- (-2.4,1);
  \draw [->,](-1.8,0) to[out=+90,in=0] (-2.4,0.5);
  \draw [dashed,] (-2.4,1) -- (3.4,1);
  \node at (-2.7,1) {$\omega$};
  \end{scope}
      \end{tikzpicture}
      \caption{Diagram showing that $-\pi/2 \leq G(j\omega) \leq \pi/2$ in accordance to condition \textbf{(2)}.}
      \label{fig:c2}
\end{figure}

Condition \textbf{(1)} is trivially verified, since the real part of both poles $s=0$ and $s=-\gamma_1-\gamma_2$ is equal to or less than zero. By inspection, condition \textbf{(2)} can be easily verified by plotting the imaginary part in the $y$-axis and the real part in the $x$-axis. This is depicted in Fig. \ref{fig:c2}, where it can be seen that, for a fixed $\omega > 0$, the condition translates into $\alpha - \beta - \pi/2 \geq -\pi/2$, which is always verified. Similarly, for a fixed $\omega < 0$, we have the symmetric case in which $-\alpha + \beta + \pi/2 \leq \pi/2$, which is always verified. Thus, $G(s)$ is positive real and system (\ref{eq:ccas2}) is absolutely stable.

\section{Numerical Simulations}\label{sec:ns}
\begin{figure} [h]
\centering
\def\svgwidth{.5\columnwidth}
\begin{tikzpicture}
  \begin{scope}[shift={(-1,0)},scale=2]
  \node 
  at (-1,0) {
      \includegraphics[width=0.5\textwidth]{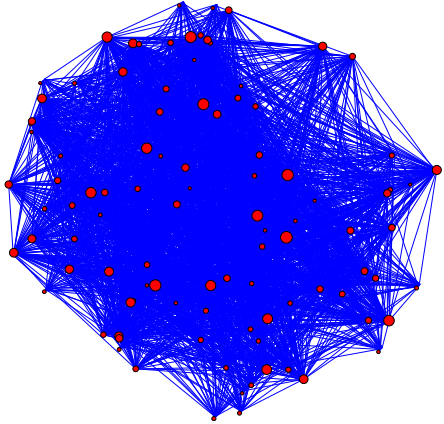}
    };
  \end{scope}
      \end{tikzpicture}
\caption{Complex network used for the simulations.}
  \label{fig:kaku}
\end{figure}

In this section we simulate the system in the case of structured environment, using the Barab\'asi-Albert complex network. We assume that only a few nodes have high connectivity, whereas a large number of nodes have very low connectivity. We use a discretized version of the following power-law distribution, see \cite{Moreno02}:
\begin{equation} \label{eq:pwlaw}
P(k) = \frac{2m^2}{k^3} \quad \mbox{for $k \geq m$}, \quad m = \langle k \rangle/2.
\end{equation}
In the rest of the section, we write $k_i=N\%$ to mean that players in class $k_i$ are connected to $N\%$ of the population. The sum of all players of all classes is in accordance with (\ref{eq:pwlaw}), i.e. $\sum_i k_i = 1$, for all $i$. The complex network is depicted in Fig. \ref{fig:kaku}.

\begin{figure}
\includegraphics[width=0.45\textwidth]{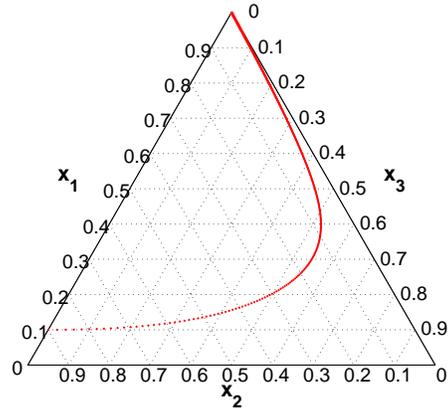}
\caption{Trajectories for $\sigma = 3$ in barycentric coordinates.}
\label{fig:3}
\end{figure}

\begin{figure}
\includegraphics[width=0.45\textwidth]{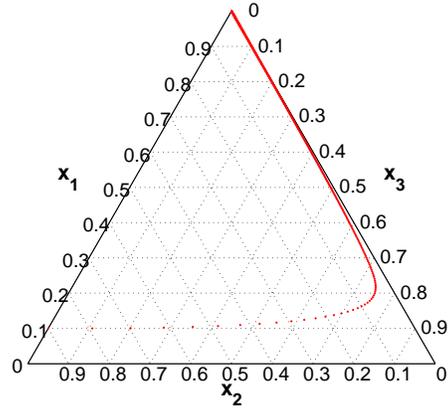}
\caption{Trajectories for $\sigma = 15$ in barycentric coordinates.}
\label{fig:4}
\end{figure}

\noindent
\textbf{The asymmetric case}. In the asymmetric case the system shares similarities with a mass-spring-damper model, as formulated in (\ref{eq:s2sys}). We investigate the role of the cross-inhibitory signal parameter $\sigma$. The plot of the population distribution is displayed in Figs. \ref{fig:3}-\ref{fig:4} for $\sigma=3$ and $\sigma=15$, respectively. The simulations involve only the population connected to only 5\% of the whole population. Such population amounts to 30\% of the total. As initial state, the population in state $1$ is equal to 10\% of the total and in state $2$ is equal to the rest 90\%. The plots show that a higher value of $\sigma$ leads to a higher transient response of the third state component and to a faster response of the first two state components. 

%

\begin{figure}
\includegraphics[width=0.45\textwidth]{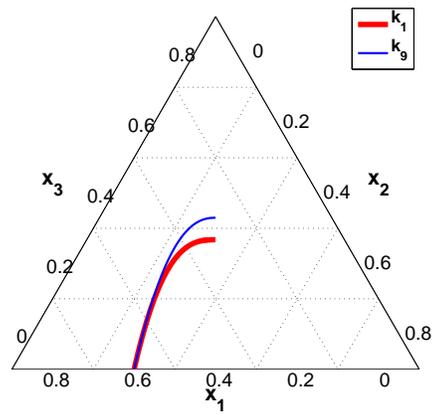}
\caption{Trajectories for $\theta_1 = \theta_2 = 0.4$ and $\sigma = 3$.}
\label{fig:7}
\end{figure}

\begin{figure}
\includegraphics[width=0.45\textwidth]{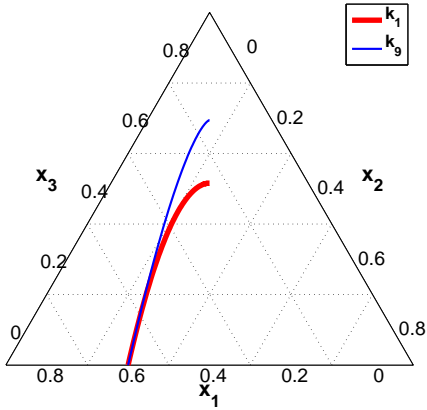}
\caption{Trajectories for $\theta_1 = \theta_2 = 0.4$ and $\sigma = 15$.}
\label{fig:8}
\end{figure}

\noindent
\textbf{Mean-Field Response}. We now simulate the mean-field response assuming a constant value $\theta_1 = \theta_2 = 0.4$, for two classes of players, namely those with connectivity $k_1 = 22\%$ and $k_9 = 85\%$. As for the initial state, the population is split among the three states as: 60\% in state $1$, 20\% in both states $2$ and $3$. From the plots in Figs. \ref{fig:7}-\ref{fig:8}, it is evident that the class with higher connectivity converges to an equilibrium point with higher values of $x_3$ (the uncommitted state). Theorem \ref{th4} justifies this behaviour, i.e. the role of parameter $\psi_k$.  

%

\begin{figure}
\includegraphics[width=0.45\textwidth]{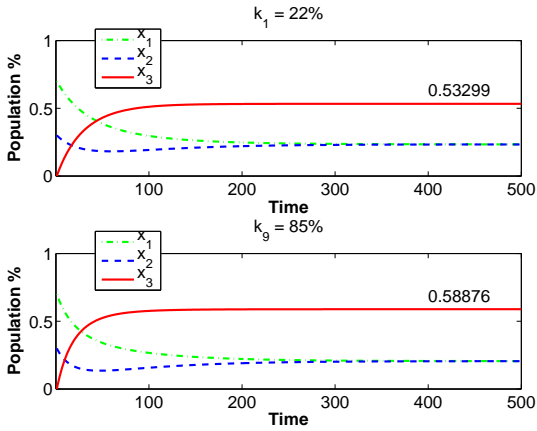}
\caption{Time evolution of micro-macro model $\sigma = 5$.}
\label{fig:9}
\end{figure}

\noindent
\textbf{Micro-macro model}. The last set of simulations involves the micro-macro model in (\ref{eq:psist11111}) and (\ref{eq:psi2}). The classes are identical to the previous set, while as for the initial state, the population is split among the three states as: 70\% in state $1$ and 30\% in state $2$. The plots in Fig. \ref{fig:9} show that at the equilibrium the value of $x_3$ increases with the connectivity, when $\sigma$ is constant, namely we have more players in the uncommitted state. Again, this is in accordance to theorem \ref{th4}. 

\section{Conclusion}\label{sec:conc}
For a collective decision making process originating in the context of honeybees swarms, we have provided an evolutionary game interpretation and we have studied stability in the case of structured and unstructured environment. Furthermore, we have investigated the role of the connectivity in terms of speed of convergence and characterisation of the equilibrium point. Finally, we have analysed the system in case of uncertain cross-inhibitory signal, which generalizes the constant coefficient used in the previous studies. 

%



\section*{Appendix}

\begin{table*}[ht]
\begin{center}
\begin{equation} \nonumber
\left[ \begin{array}{c}
\dot{\theta_1} \\
\dot{\theta_2} \end{array} \right] = 
\left[ \begin{array}{cc}
\frac{r}{k_{max}}(V(k)/\langle k \rangle -\Psi_1 - \Psi_2) -\alpha - \gamma & -\frac{\sigma}{k_{max}}\Psi_1 -\gamma\\
-\frac{\sigma}{k_{max}}\Psi_2 -\gamma & \frac{r}{k_{max}}(V(k)/\langle k \rangle -\Psi_1 - \Psi_2) -\alpha - \gamma \end{array} \right]
\left[ \begin{array}{c}
\theta_1 \\
\theta_2 \end{array} \right] +
\left[ \begin{array}{c}
\gamma  \\
\gamma \end{array} \right].
\end{equation}
\end{center}
\caption{System \eqref{eq:psi2} in matrix form.}
\label{Jacobian}
\end{table*}

\noindent
\textbf{Proof of Theorem \ref{th1}}.
To study equilibrium points, we first impose $\dot{x}_1 = \dot{x}_2 = 0$ and obtain $(x_3r - \alpha)(x_1-x_2) = 0$, which leads to two solutions: $x_1 = x_2$ and $x_3 = \alpha/r$, both studied in the first two cases, while the third case analyses the scenario in which both hold true.  

[\textbf{Case 1}] When $x_1 = x_2 = x$, the equilibrium point is the root of a second degree polynomial, $\dot{x}  = (2r+\sigma) x^2 - (r - 2\gamma - \alpha) x - \gamma = 0$.
 From $\dot x_1=0$ and  $x_1^*+x_2^*+x_3^*=1$, we have the following equilibrium point, the roots of the above  polynomial are given by:
\begin{equation}\label{eq:eqcase1.2}\nonumber
\begin{array}{lll}
x_1^*  = \frac{(r - 2\gamma - \alpha) + \sqrt{(r - 2\gamma -\alpha)^2+4\gamma(2r+\sigma)}}{2(2r + \sigma)} = x_2^*, \\
x_3^*=1- \frac{(r - 2\gamma - \alpha) + \sqrt{(r - 2\gamma -\alpha)^2+4\gamma(2r+\sigma)}}{2(2r + \sigma)}.
\end{array}
\end{equation}

\noindent 
[\textbf{Case 2}] When $x_3 = \alpha/r$, setting $\dot{x}_1 = 0$ and replacing $x_3 = \alpha/r$  in (\ref{eq:gfs}), we obtain $\dot{x}_1=\alpha x_1 + \frac{\alpha}{r}\gamma - \alpha x_1+\sigma x_1 -\sigma x_1^2 - \frac{\alpha}{r} \sigma x_1 = 0$,
which in turn implies $x_1^2 - (1-\frac{\alpha}{r})x_1 - \frac{\alpha \gamma}{r \sigma} = 0.$
From the roots of the above polynomial, the two equilibrium points are:
\begin{equation} \label{eq:eqcase2.4}
\begin{array}{lll}
x_1^* = \frac{1-\frac{\alpha}{r} \pm \sqrt{(1-\frac{\alpha}{r})^2 + \frac{4 \alpha \gamma}{\sigma r}}}{2}, \\
x_2^* = 1- \frac{1-\frac{\alpha}{r} \pm \sqrt{(1-\frac{\alpha}{r})^2 + \frac{4 \alpha \gamma}{\sigma r}}}{2}- \frac{\alpha}{r}, \quad x_3^* = \frac{\alpha}{r}.
\end{array}
\end{equation}

\noindent 
[\textbf{Case 3}] A special case is when $x_3=\alpha/r$ and $x_1=x_2$. $\dot{x}  =  \frac{\alpha}{r}(rx+\gamma) - x(\alpha+\sigma x)  =  \alpha x + \frac{\gamma \alpha}{r} - \alpha x - \sigma x^2 =  \sigma x^2 - \frac{\gamma \alpha}{r} = 0$.
Since $x_1 = x_2$, this leads to:
$x_1=x_2 = \sqrt{\frac{\alpha \gamma}{r \sigma}},$
which is equivalent to:
$x_1  = \frac{1- \alpha/r}{2} = \frac{r-\alpha}{2r}.$
From the previous equations, we obtain the following value for $\sigma$:
\begin{equation}\label{eq:eqcase3.4}
\frac{(r-\alpha)^2}{4r^2} = \frac{\alpha \gamma}{r \sigma} \quad \Rightarrow \quad
\sigma = \frac{4 r \alpha \gamma}{(r-\alpha)^2}.
\end{equation}
Thus we have the equilibrium point
\begin{equation} \label{eq:eqcase3.5}\nonumber
\begin{array}{lll}x^*=  \Big(
\sqrt{\frac{\alpha \gamma}{r \sigma}}, 
\sqrt{\frac{\alpha \gamma}{r \sigma}}, \frac{\alpha}{r}\Big)
=  \Big(
\frac{r-\alpha}{2r}, 
\frac{r-\alpha}{2r}, \frac{\alpha}{r}\Big).
\end{array}
\end{equation}

\noindent
\textbf{Proof of Theorem \ref{th2}}.
From $x_3=1-x_1-x_2$, let us rewrite (\ref{eq:gfs}) as 
\begin{equation} \label{eq:gfsrew}
\left\{\begin{array}{lll}
\dot{x}_1 = (1-x_1-x_2) (rx_1+\gamma) - x_1(\alpha+\sigma x_2), \\
\dot{x}_2 = (1-x_1-x_2)(rx_2+\gamma) - x_2(\alpha+\sigma x_1). \\
\end{array}\right.
\end{equation}
To analyse the stability of system (\ref{eq:gfsrew}), we compute the Jacobian matrix 
 around an equilibrium point, i.e. $x:=x_1 = x_2$, as 
\begin{equation} \label{eq:jac2}
\left[ \begin{array}{cc}
r - 3rx - \gamma -\alpha -\sigma x & x(-r - \sigma) - \gamma  \\
x(-r - \sigma) - \gamma & r - 3rx - \gamma -\alpha -\sigma x \end{array} \right],
\end{equation}
for which we have a saddle point  when the following condition for the determinant $\Delta$ of the Jacobian holds: $\Delta:=J_{11}J_{22}-J_{12}J_{21}=J_{11}^2-J_{12}^2< 0.$ The latter is true when $x(r + \sigma) + \gamma > 3rx + \gamma +\alpha +\sigma x - r,$
which in turn implies $-2rx > \alpha - r.$
The latter yields $x < \frac{r-\alpha}{2r}$.
From considering $x=\sqrt{\frac{\alpha \gamma}{r\sigma} }$ for the equilibrium in \textbf{Case 3}, it follows
\begin{equation} \label{eq:sigma3}
\sigma > \frac{4r\alpha \gamma}{(r-\alpha)^2}.
\end{equation}
This concludes our proof.

\noindent
\textbf{Proof of Theorem \ref{th3}}.
The determinant $\Delta$ of the above matrix is always positive. To see this, note that $ (r + \sigma)  \psi_k  \theta  + \alpha + \gamma  \geq  
 \psi_k r \theta + \gamma.$
Also, the trace of the above matrix is negative, i.e., $T=- 2 (r + \sigma)  \psi_k  \theta  - 2\alpha - 2\gamma <0,$ and therefore the system is asymptotically stable. From $T^2-4\Delta = 4 (\psi_k r \theta + \gamma)^2 >0$ we can conclude that the equilibrium point is an asymptotically stable node. \\ 
As for the speed of convergence, let us focus on the eigenvalues of the Jacobian. To this purpose, let us calculate the determinant which is given by
\begin{equation}\label{eq:delta}\nonumber
\begin{array}{ll}
\Delta =  [-(\sigma + r) \psi_k \theta -\alpha -\gamma]^2 - [-\psi_k r \theta - \gamma]^2 \\
 = \sigma^2 \psi_k^2 \theta^2 +\alpha^2 -2[ r\sigma \psi_k^2 \theta^2 - r \alpha \psi_k \theta -\sigma \alpha \psi_k \theta -\sigma \gamma \psi_k 
 - \alpha \gamma] \\
 = (\sigma^2 - 2r\sigma)\psi_k^2 \theta^2 + (2r\alpha + 2\sigma\alpha + 2\sigma\gamma)\psi_k\theta +\alpha^2 +2\alpha\gamma.
\end{array}
\end{equation}
Then, $T^2 - 4\Delta = 4[\psi_k r\theta + \gamma]^2.$
Thus, the eigenvalues of the Jacobian matrix are $\lambda_{1,2} = -(\sigma+ r) \psi_k\theta -\alpha -\gamma \pm (\psi_k r \theta + \gamma).$
In the two extreme case of no connectivity $\psi_k = 0$ and full connectivity $\psi_k = 1$. 
\begin{equation}\label{eq:labdas}\nonumber
\lambda_{1,2} = \left\{ \begin{array}{ll}
 (-\alpha -2\gamma, -\alpha), \, for \, \psi_k = 0, \\
 (-(2r + \sigma)\theta -\alpha -2\gamma, -\sigma\theta -\alpha), \, for \, \psi_k = 1.
\end{array} \right.
\end{equation}

\noindent
\textbf{Proof of Theorem \ref{th4}}.
We can compute the following
\begin{equation}\label{eq:eqmf}
\begin{array}{ll}
x_k^* =  A^{-1}_k(\theta)  c_k(\theta)  \\
\quad = \frac{1}{-(2r + \sigma)\psi_k\theta-\alpha-2\gamma} [-\psi_kr\theta - \gamma \quad -\psi_kr\theta - \gamma]^T \\
\quad = \frac{1}{(2r + \sigma)\psi_k\theta+\alpha+2\gamma} [\psi_kr\theta + \gamma \quad \psi_kr\theta + \gamma]^T.
\end{array}
\end{equation}
Again, when considering the above two cases we get
\begin{equation}\label{eq:eqmf2}
\begin{array}{ll}
x_k^* =  \frac{1}{\alpha+2\gamma} [\gamma \quad \gamma]^T \quad \psi_k =0, \\
x_k^* =  \frac{1}{(2r + \sigma)\theta+\alpha+2\gamma} [r\theta + \gamma \quad r\theta + \gamma]^T \quad \psi_k =1.
\end{array}
\end{equation}
Therefore, we can also say that higher connectivity increases the number of players in the uncommitted state.

\noindent
\textbf{Proof of Theorem \ref{thsig}}.
To compute the equilibrium, let us set $\dot{\theta}_1 = \dot{\theta}_2$ and obtain:
\begin{equation} \label{eq:psi3}
\Big(\theta_1 - \theta_2\Big)\Big(\frac{r}{k_{max}}\Psi_3 - \alpha\Big) + \frac{\sigma \theta_1}{k_{max}}\Psi_2 - \frac{\sigma \theta_2}{k_{max}}\Psi_1 = 0.
\end{equation}
Note that in a symmetric equilibrium where $\theta_1 = \theta_2$, we can neglect the last two terms. We can then compute the Jacobian of system \eqref{eq:psi2}  and obtain the matrix in Table \ref{Jacobian}.

To inspect the existence of saddle points we need to study conditions under which the determinant of the Jacobian is less than 0. Then, we take $\Psi_1 = \Psi_2$ and impose that the right-hand side is greater than the left-hand side
\begin{equation*}
\Big(-\frac{\sigma}{k_{max}}\Psi - \gamma\Big)^2 > \Big(\frac{r}{k_{max}}(V(k)/\langle k \rangle-2\Psi) -\alpha -\gamma\Big)^2.
\end{equation*}
By taking the square root on both sides, since the left-hand side is strictly negative, we have  $-\frac{\sigma}{k_{max}}\Psi < -2\frac{r}{k_{max}}\Psi + \frac{rV(k)}{k_{max}\langle k \rangle} -\alpha$,
and after some basic algebra, we get (\ref{eq:psi5}).

\noindent
\textbf{Proof of theorem \ref{th5}}.
Let us first prove that $Z(s)$ is strictly positive real. Thus, we study the properties of matrix $Z(s)$, specifically the positive realness. To be strictly positive real, the following conditions must hold true:
\begin{itemize}
\item $Z(s)$ is Hurwitz, i.e poles of all elements of $Z(s)$ have negative real parts;
\item $ Z(j\omega) + Z(-j\omega) > 0, \quad \forall \omega \in \mathbb{R};$
\item $Z(\infty) + Z^T(\infty) >0$.
\end{itemize}
First, we prove that $Z(s)$ is Hurwitz. Thus, all the poles must be negative, i.e. $r - 4rx-2\gamma-\alpha<0$. which holds true, after considering the discussion on the trace of matrix $A$ as a direct consequence. Now, we check the second condition. It follows that
\begin{equation} \label{eq:cccscZ}\nonumber
\begin{scriptsize}
\begin{array}{lll}
Z(j\omega) + Z(-j\omega) = \frac{1}{j\omega+\zeta} \left[ \begin{array}{cc}
j\omega+\zeta + k & k \\
k & j\omega+\zeta + k  \end{array} \right] \\+ \frac{1}{-j\omega+\zeta} \left[ \begin{array}{cc}
-j\omega+\zeta + k & k \\
k & -j\omega+\zeta + k  \end{array} \right] = 
\left[ \begin{array}{cc}
\frac{j\omega+\zeta + k}{j\omega+\zeta} & \frac{k}{j\omega+\zeta} \\
\frac{k}{j\omega+\zeta} & \frac{j\omega+\zeta + k}{j\omega+\zeta}  \end{array} \right] \\
+ \left[ \begin{array}{cc}
\frac{-j\omega+\zeta + k}{-j\omega+\zeta} & \frac{k}{-j\omega+\zeta} \\
\frac{k}{-j\omega+\zeta} & \frac{-j\omega+\zeta + k}{-j\omega+\zeta}  \end{array} \right]  =
\left[ \begin{array}{cc}
z_{11} & z_{12} \\
z_{21} & z_{22}  \end{array} \right],
\end{array}
\end{scriptsize}
\end{equation}
where \\$z_{11} = z_{22} = \frac{\omega^2-j\omega\zeta - j\omega k +j\omega\zeta +\zeta^2+\zeta k +\omega^2+j\omega\zeta +j\omega k -j\omega\zeta+\zeta^2+\zeta k}{\zeta^2+\omega^2}$ and $z_{12} = z_{21} = \frac{-j\omega k+\zeta + k + j\omega k +\zeta k}{\zeta^2+\omega^2}$. Thus, the second condition can be rewritten as
\begin{equation} \label{eq:cccscZ}\nonumber
\begin{scriptsize}
\begin{array}{lll}
Z(j\omega) + Z(-j\omega) =
\left[ \begin{array}{cc}
\frac{2\omega^2+2\zeta^2+2\zeta k}{\zeta^2+\omega^2} & \frac{2\zeta + k}{\zeta^2+\omega^2} \\
\frac{2\zeta + k}{\zeta^2+\omega^2} & \frac{2\omega^2+2\zeta^2+2\zeta k}{\zeta^2+\omega^2} \end{array} \right]>0,
\end{array}
\end{scriptsize}
\end{equation}
which is verified for all $\omega$. Last, as  $Z(s)$ is symmetric the third condition implies that $2Z(\infty) > 0$ . Note that the off-diagonal entries converge to zero in the limit, while entries on the main diagonal converge to 1 in the limit. So we have an identity matrix, and thus the third condition is verified. Let us turn to prove absolute stability by showing that there exists a Lyapunov function $V(x) = x^TPx$. Let us derive the expression of $\dot{V}(t,x)$ as
\begin{equation} \label{eq:ccVdot}
\begin{array}{lll}
\dot{V}(t,x) & = \dot{x}^TPx + x^TPx \\
& = x^TA^TPx + x^TPAx - \psi^TB^TPx - x^TPB\psi,
\end{array}
\end{equation}
where $\psi$ is equivalent of writing $\psi(t,y)$. For the condition on the sector nonlinearity $-2\psi^T(\psi- Ky) \ge 0$ and from matrices $P$ and $K$ being symmetric, we can now specialize it to our case, i.e. $A$ symmetric and $B = C = \mathbb{I}$, as
\begin{equation} \label{eq:ccVdot2}
\begin{array}{lll}
\dot{V}(t,x) & \le  x^T(A^TP + PA)x - 2x^TPB\psi -2\psi^T(\psi -Ky) \\
& = 2x^TAPx - 2x^TP\psi +2\psi^TKx - 2\psi^T\psi \\
& =  2x^TAPx + 2x^T(K-P)\psi - 2\psi^T\psi.
\end{array}
\end{equation}
We can rewrite this in matrix form as
\begin{equation} \label{eq:Vdot3}
\begin{array}{lll}
\dot{V}(t,x) \le 2x^T \left[ \begin{array}{cc}
a & b \\
b & a  \end{array} \right]
\left[ \begin{array}{cc}
p_1 & p_2 \\
p_2 & p_1  \end{array} \right] x + 2x^T 
\Bigg(\left[ \begin{array}{cc}
\tilde k & \tilde k \\
\tilde k & \tilde k  \end{array} \right] \\ \qquad -
\left[ \begin{array}{cc}
p_1 & p_2 \\
p_2 & p_1  \end{array} \right]\Bigg) x - 2\psi^T\psi,
\end{array}
\end{equation}
where $a = r-3rx-\gamma-\alpha$ and $b = -rx-\gamma$. To show that the right-hand side of (\ref{eq:ccVdot2}) is negative, we can construct a square term by imposing
\begin{equation} \label{eq:ccPL}
\begin{array}{lll}
2AP & =   - L^TL -\epsilon P, \\
K - P & =  \sqrt{2}L^T,
\end{array}
\end{equation}
where $\epsilon>0$ is a constant and matrix $P = P^T >0$. Now, we can rewrite (\ref{eq:ccVdot2}) as
\begin{equation} \label{eq:ccVdot3}
\begin{array}{lll}
\dot{V}(t,x) & \le -\epsilon x^TPx - x^TL^TLx + 2\sqrt{2}x^TL^T\psi - 2\psi^T\psi \\
& = -\epsilon x^TPx - [Lx - \sqrt{2}\psi]^T[Lx - \sqrt{2}\psi] \\
& \le -\epsilon x^TPx.
\end{array}
\end{equation}
From Kalman-Yakubovich-Popov lemma, we can obtain $P$, $L$, $\epsilon$ solving (\ref{eq:ccPL}), as $Z(s)$ is positive real. This concludes our proof.

\end{document}